\documentclass[prl,twocolumn,showpacs,superscriptaddress,floatfix]{revtex4}
\usepackage{graphicx}

\usepackage{amsmath}
\usepackage{amssymb}
\usepackage{amsbsy}
\usepackage{textcomp}

\usepackage{color}
\definecolor{orange}{rgb}{1,0.5,0}

\begin{document}

\title{Instanton filtering for the stochastic Burgers equation}

\author{Tobias \surname{Grafke}}
\affiliation{Theoretische Physik I, Ruhr-Universit\"at Bochum,
Universit\"atsstr. 150, D44780 Bochum (Germany)}
\author{Rainer \surname{Grauer}}
\affiliation{Theoretische Physik I, Ruhr-Universit\"at Bochum,
Universit\"atsstr. 150, D44780 Bochum (Germany)}
\author{Tobias Sch\"afer}
\affiliation{Department of Mathematics, College of Staten Island, CUNY, USA}

\date{\today}

\begin{abstract}
  We address the question whether one can identify instantons in
  direct numerical simulations of the stochastically driven Burgers
  equation. For this purpose, we first solve the instanton equations
  using the Chernykh-Stepanov method [Phys. Rev. E \textbf{64}, 026306
  (2001)]. These results are then compared to direct numerical
  simulations by introducing a filtering technique to extract
  prescribed rare events from massive data sets of realizations. Using
  this approach we can extract the entire time history of the
  instanton evolution which allows us to identify the different phases
  predicted by the direct method of Chernykh and Stepanov with
  remarkable agreement.
\end{abstract}

\maketitle

\noindent {\it Introduction}
Understanding intermittency in turbulent flows is still one of the
open problems in classical physics. More than 15 years ago, for
certain systems like the problem of passive advection and Burgers
turbulence the door for attacking this issue was opened by getting
access to the probability density function to rare and strong
fluctuations by the instanton approach
\cite{shraiman-siggia:1994,gurarie-migdal:1996,falkovich-kolokolov-etal:1996,balkovsky-falkovich-etal:1997}.
In this letter we concentrate on rare fluctuations in Burgers
turbulence. For that case, Gurarie and Migdal
\cite{gurarie-migdal:1996} introduced the instanton approach and were
able to calculate the instanton contribution to the right tail of the
velocity increment probability distribution function (PDF). In
succeeding work, \citet{balkovsky-falkovich-etal:1997}
were able to characterize the left tail of the increment PDF making
use of the Cole-Hopf transformation \cite{hopf:1950,cole:1951}. These
analytical results were confirmed by direct numerical solution of the
instanton equations by \citet{chernykh-stepanov:2001}.

The open question remained whether one can observe or identify the instanton in
numerical simulations of the stochastic Burgers equation. The answer is not
obvious, since one could argue that, perhaps, the contribution of the instanton is
exponentially small such that instantons are only relevant to such rare events
that they are not interesting from a practical point of view at finite Reynolds
numbers. In this Letter, however, we find that already at moderate Reynolds
numbers the instanton can be identified in data sets of simulations of the
stochastic Burgers equation. This gives a positive answer to this important
question. In particular, we show by introducing a particular filtering technique
that all phases of the instanton evolution can be recovered from data sets of
simulations of the stochastic Burgers equation.

The Letter is organized as follows: we first review the path integral
formulation for Burgers turbulence and revisit the algorithm
introduced by Chernykh and Stepanov to solve directly the instanton
equations. We then describe our numerical simulations to obtain
sufficient statistical data ($\approx 10^7$ realizations of complete
stochastic Burgers simulations using CUDA graphics cards) necessary
for our instanton filtering. Using this enormous data set we apply our
instanton filtering procedure and compare the results with the direct
instanton simulations. A conclusion and outlook summarize the
Letter.\\

\noindent{\it Action functional and instanton equations}. 
We consider the stochastically driven Burgers equation given by
\begin{equation} \label{stochastic_burgers}
u_t+uu_x-\nu u_{xx} = \phi
\end{equation}
with a noise field $\phi$ that is $\delta$-correlated in time and has
finite correlation in space with correlation length $L$, more
precisely
\begin{eqnarray}
\langle\phi(x,t)\phi(x',t')\rangle &=& \delta(t-t')\chi((x-x')/L), \\
\chi(x)& =& (1-x^2){\mathrm{e}}^{-x^2/2}\,\,.
\end{eqnarray}
While the precise form of $\chi$ is not important for the results of our work,
we chose this particular form in order to have the same setup of the problem as
in previous studies by Chernykh and Stepanov. Using the functional path integral
introduced by Martin-Siggia-Rose/Janssen-de Dominicis
\cite{martin-siggia-rose:1973,janssen:1976,dedominicis:1976,phythian:1977}, the
PDF of the velocity gradients $u_x(t=0,x=0)$ is written as
\begin{equation}
{\mathcal{P}}(a) = \int {\mathcal{D}}u\,{\mathcal{D}}p\,d{\mathcal{F}}\, \exp\left(-\tilde S(u,p,{\mathcal{F}})\right)\,\,,
\end{equation}
with the action $\tilde S$ given by
\begin{eqnarray}
\tilde S &=& \frac{1}{2} \int _{-\infty}^0 dt \int dx_1\,dx_2 \,\,p(x_1,t)\chi(x_1-x_2)p(x_2,t) \nonumber \\
&& -i\int_{-\infty}^0 dt
\int dx\,\, p(u_t+uu_x-\nu u_{xx}) \nonumber \\ && -4\nu^2{\mathcal{F}}i(u_x(0,0)-a) \;\; ,
\end{eqnarray}
where $\mathcal{F}$ results from the Fourier transform of the $\delta$-function
for the observable $u_x(t=0,x=0) = a$.
The saddle point (instanton) equations for the fields ($u$,$p$)
yielding the largest contribution to the path integral for strong
gradients are then given by
\begin{subequations}
 \label{instanton}
\begin{eqnarray}
u_t + uu_x - \nu u_{xx} &=& -i \int \chi(x-x')p(x',t)dx' \label{instantonu}\\
p_t + up_x + \nu p_{xx} &=& 4i\nu^2{\mathcal{F}} \delta(t)\delta'(x) \;\; . \label{instantonp}
\end{eqnarray}
\end{subequations}

\noindent  {\it The Chernykh-Stepanov algorithm revisited}.
The algorithm proposed by Chernykh and Stepanov for solving the
system of partial differential equations for the fields $u$ and $p$ 
can be summarized as follows: the diffusion terms in
the equations (\ref{instanton}) define the temporal direction of the
numerical integration of the equations, meaning that $u$ is integrated
forward in time while $p$ is integrated backwards. The
right-hand side of equation (\ref{instantonp}) poses the initial
condition $p(t=0,x) = -4i\nu^2{\mathcal{F}} \delta'(x)$ and the
starting step is obtained by setting $u(t,x)=0$. Equation
(\ref{instantonp}) is then solved backward in time up to a large
negative time mimicking $-\infty$. The obtained solution $p(t,x)$ is
used in the right-hand side of equation (\ref{instantonu}) such that
this equation can be solved forward in time. This procedure is then
iterated until convergence. For higher gradients, a stabilization of
this iteration has to be applied, details of which can be found in
\cite{chernykh-stepanov:2001}. While Chernykh and Stepanov use a
stabilized finite difference scheme with an implicit first-order time
integration, we utilize a second-order Adams-Bashforth temporal
integration for a pseudo-spectral method. We also note the similarity
of the system (\ref{instanton}) to equations that arise in the context
of transition probabilities
\cite{e-ren-etal:2004,fogedby-ren:2009}. Although the boundary
conditions are different, the above system of instanton equations can,
in principle, also be solved numerically by minimizing the
corresponding action using a L-BFGS scheme. We found, however, for the
case under consideration, the propagation-based Chernykh-Stepanov
scheme numerically much more efficient.  Therefore, the
Chernykh-Stepanov scheme might be an interesting alternative to
compute transition probabilities. A detailed comparison of both
schemes is
beyond the particular scope of this paper and will presented elsewhere. \\

\noindent {\it Parallel simulation of the stochastic Burgers equation.}
In order to generate data from simulations of the stochastic Burgers
equation, we need to solve eq. (\ref{stochastic_burgers}) with the
appropriate right-hand side. For the generation of the stochastic
force field $\phi$, at each step in time, we draw a vector $r$ of
appropriately scaled normally distributed random numbers. The size of
the vector corresponds to the discretization in $x$. This vector $r$
is then multiplied by a matrix $A$ resulting from the Cholesky
decomposition of the (discretized) correlation matrix $C$. Note that
naive discretization of $\chi$ may lead to a $\tilde C$ that, due to
finite machine-precision, is not positive-semidefinite. Therefore we
used the algorithm introduced in \cite{qi-sun:2006} in order to obtain
a matrix $C$ that is positive-semidefinite and sufficiently close to
$\tilde C$. Note that this method of generating the noise term in
eq. (\ref{stochastic_burgers}) is different from Fourier-based methods
that are commonly used for such simulations
\cite{cheklov-yakhot:1995b,gotoh-1999} and more computationally
expensive. The reason for the presented choice was motivated by the
necessity to numerically generate noise that closely imitates the
fluctuations assumed in the instanton analysis.

As the size of an individual realization is small enough to fit on a
single graphics card with its complete history, the whole simulation
is performed in CUDA alone. Matrix-Vector-operations are realized
using the cuBLAS-package, the fast Fourier-transform is provided by
cuFFT. Since the filtering and shifting procedure was also implemented
in CUDA, a whole bulk of simulations is performed and filtered
independently on the GPU. Averaging over different CUDA-processes
occurs after completion of a bulk and is performed via MPI. Thus, both
expensive device-to-host copies and high-latency network communication
are minimized. Because of the stochastic independence of realizations,
this method scales linearly with the number of graphics cards.

\begin{figure}[t]
\begin{center}
\includegraphics[width=\columnwidth]{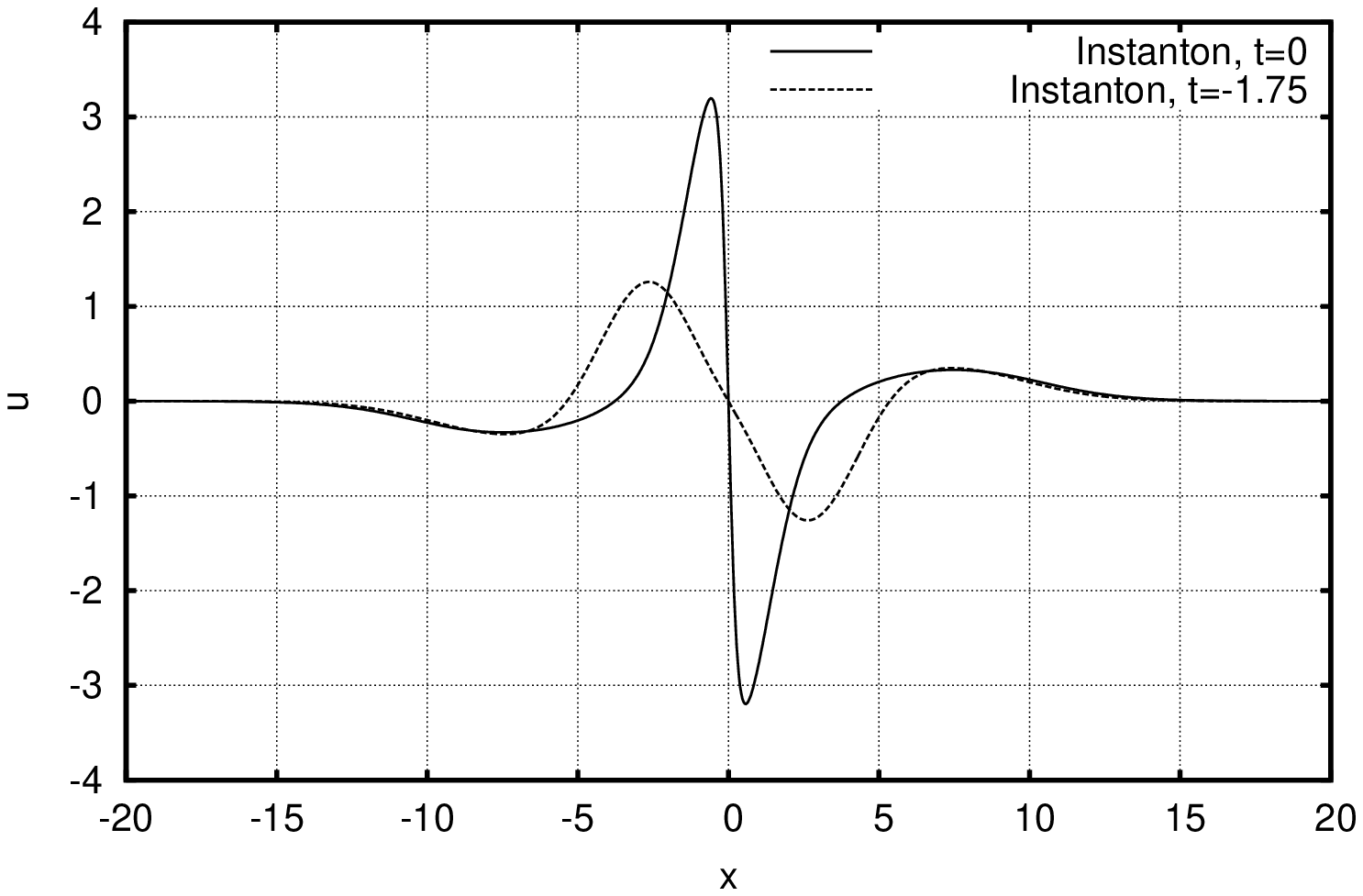} \\
\includegraphics[width=\columnwidth]{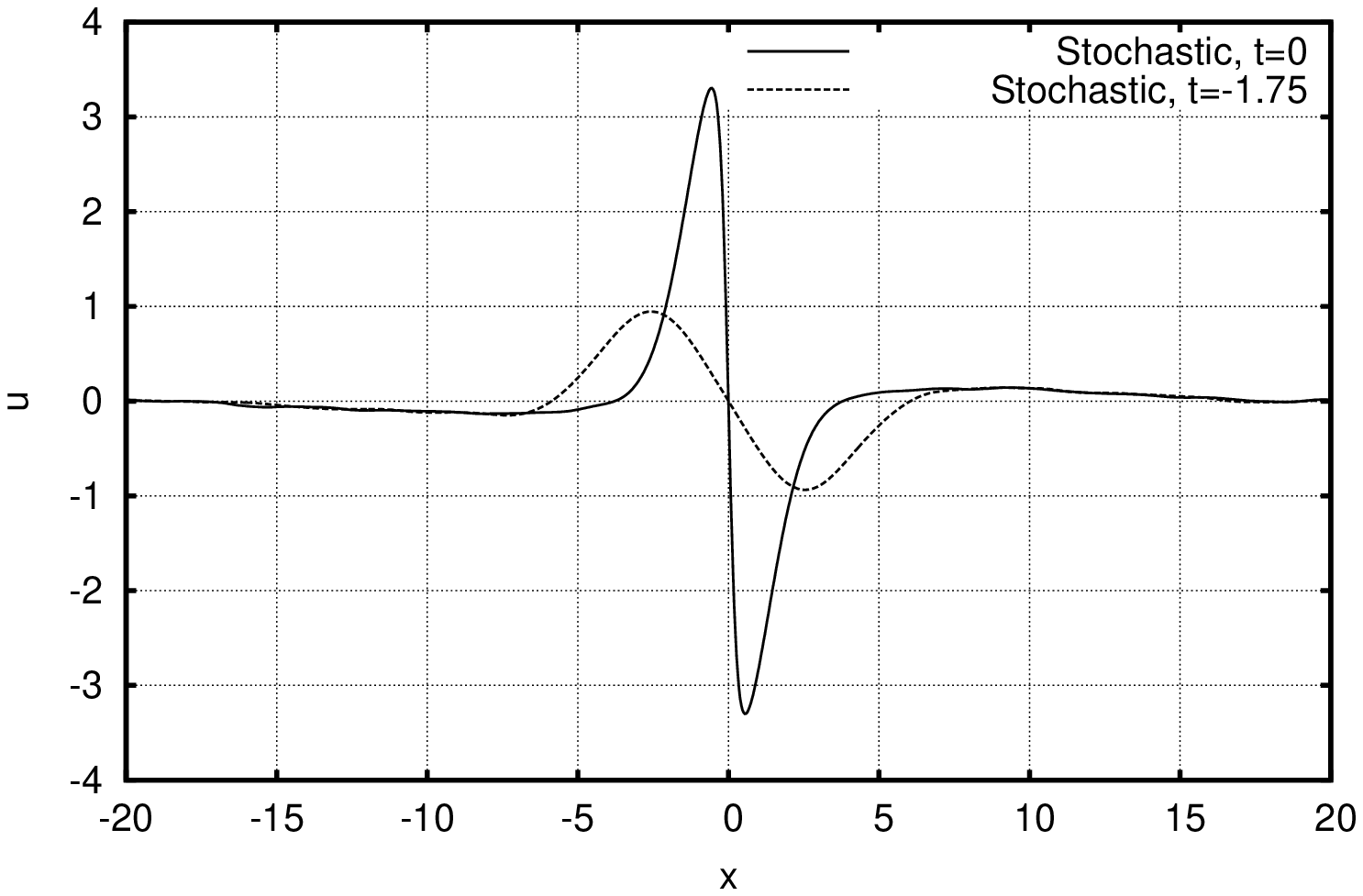}
\end{center}
\caption{Comparison of the filtered velocity field $\langle
  u_{\text{shifted}}(t,x) \rangle$ (top) and the instanton field
  $u(t,x)$ (bottom) at times $t=0$ (solid) and $t=-1.75$ (dashed).
\label{fig:comparisonInstantons}}
\end{figure}
\noindent {\it Extracting the instanton.}
In order to provide a sufficient data set for the extraction of the instanton
from simulations of the stochastic Burgers equation, we conducted the following
numerical experiment: We started the integration of the stochastic Burgers
equation from zero initial conditions for the velocity field
$u(t=t_{\text{min}},x) = 0$ at a large negative time
$t_{\text{min}}$ up to the final time $t=0$. Typical
parameters are summarized in table~\ref{table1}.
\begin{table}
  \centering
   \begin{tabular}{c|ccccccccc}
          &$N$   &$dx$ &$\eta$ &$L$ &$L_\text{box}$ &$\nu$ &$\epsilon_\mathrm{k}$&$T_L$  &$\#$hits ($\%$)  \\ \hline
     Run 1& 1024 &0.039&0.406  &1   &40             &0.3   &4.586                &0.99   & 10.5            \\
     Run 2& 1024 &0.039&0.464  &1   &40             &0.38  &2.691                &0.97   & 0.410           \\
     Run 3& 1024 &0.039&0.481  &1   &40             &0.41  &2.33                 &0.95   & 0.052           \\
   \end{tabular}
   \caption{\label{table1} Parameters of the numerical simulations.
     $N$: number of collocation points,
     $dx$: grid-spacing, 
     $\eta =(\nu^3/\epsilon_\mathrm{k})^{1/4}$: Kolmogorov dissipation length scale,
	 $L$: correlation length of forcing,
     $L_\text{box}$: domain length,
     $\nu$: kinematic viscosity, 
     $\epsilon_\mathrm{k}$: mean kinetic energy dissipation rate, 
     $T_L = L/u_\mathrm{rms}$: large-eddy turnover time,
     $\#$hits ($\%$): percentage of hits with prescribed velocity derivative.}
\end{table}
\begin{figure}[t]
\begin{center}
\includegraphics[width=\columnwidth]{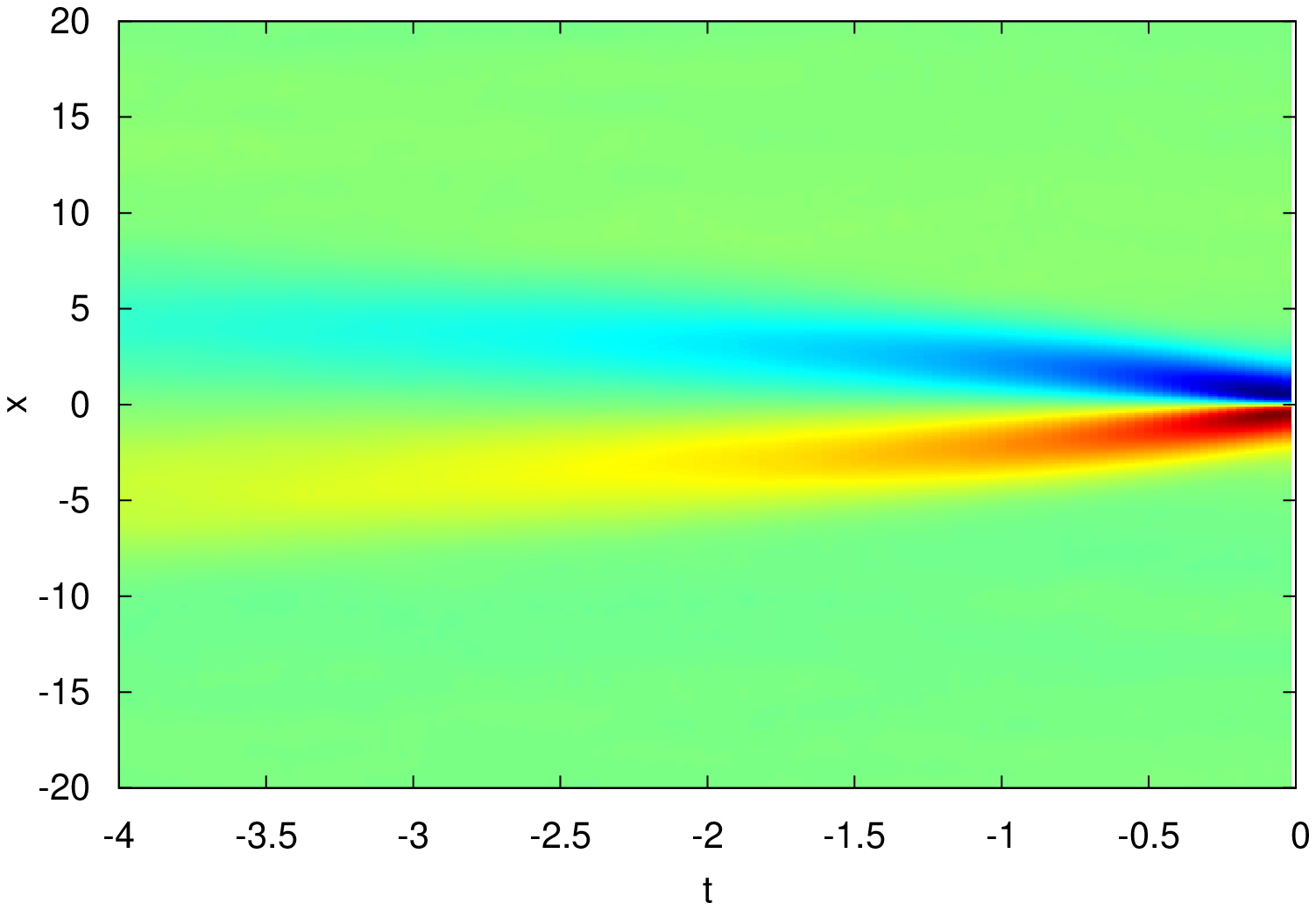} \\
\includegraphics[width=\columnwidth]{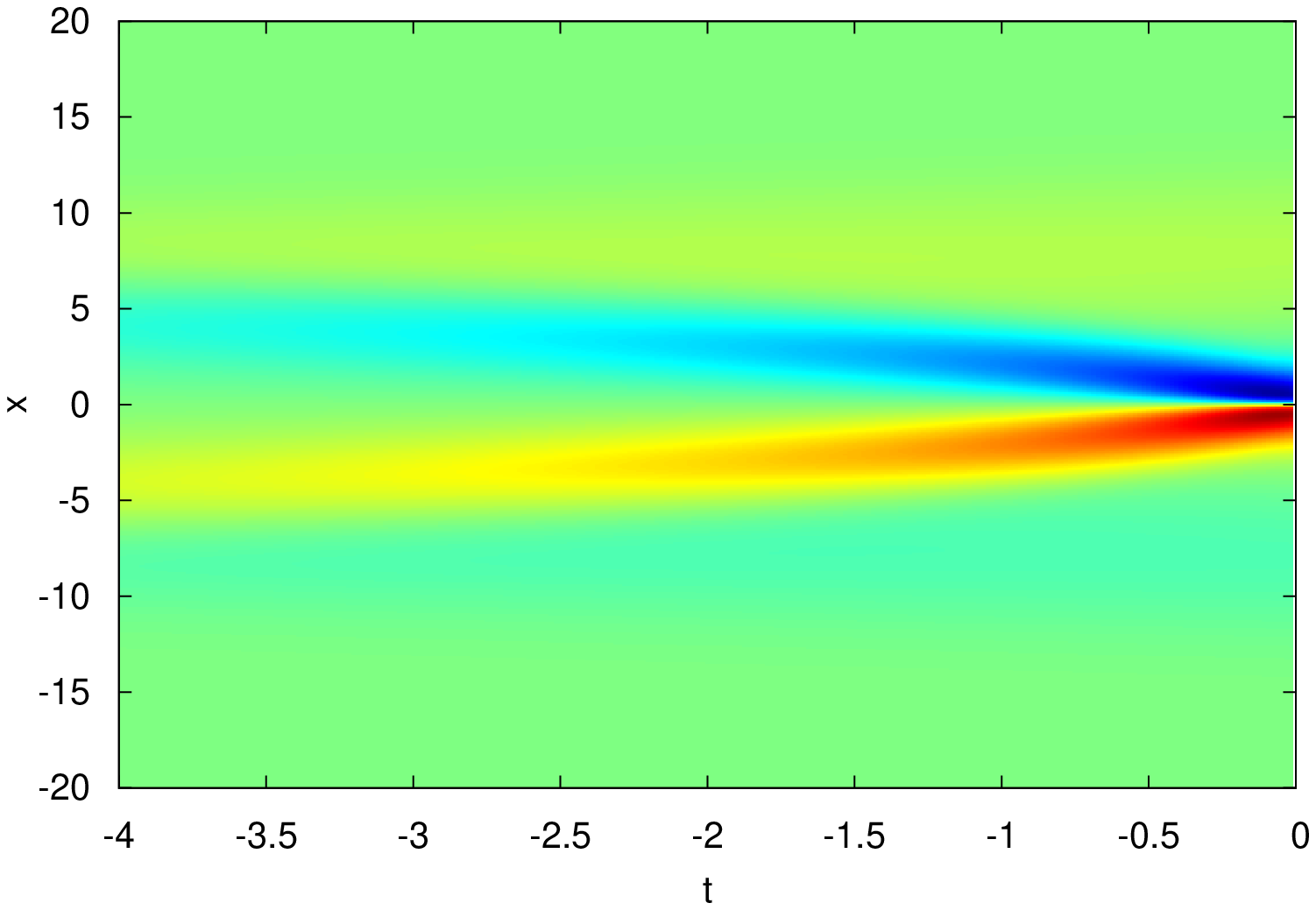}
\end{center}
\caption{Comparison of the filtered velocity field $\langle
  u_{\text{shifted}}(t,x) \rangle$ (top) and the instanton field
  $u(t,x)$ (bottom) as a space-time contour plot.
\label{fig:comparisonInstantonsContour}}
\end{figure}
The initial time $t_{\text{min}}$ was chosen corresponding to the
instanton equations (\ref{instanton}) and consists of more than $10$
integral times $T_L$. This single experiment was repeated $\approx
10^7$ times using the 64 CUDA Tesla 1060 graphics on the Bochum GPU
Cluster and the 96 CUDA Fermi 2050 graphics cards on the CUNY
GPU Cluster. The total simulation length obtained by this
parallelism corresponds to $\approx 10^8$ integral times $T_L$. 
Each of these simulations was analyzed in the following way: We
prescribed a small interval around a given value of the velocity
gradient $u_x(t=0,x)=a$ at the final time $t=0$ and searched for the
maximum velocity gradient in the numerical solution at that time. If
we find that the maximum velocity gradient falls into the desired
interval, we shift the field in space such that the location of the maximum
velocity gradient is located at $x=0$. In addition, we also shift the
forcing field $\phi(t,x)$ in the same way. The averaging procedure now
consists of taking the average of all those shifted fields
$u_{\text{shifted}}(t,x)$ and $\phi_{\text{shifted}}(t,x)$. We thus
obtain an ensemble average $\langle
u_{\text{shifted}}(t,x)\rangle$ and $\langle
\phi_{\text{shifted}}(t,x)\rangle$ in space and time. Since the
forcing field is $\delta$-correlated in time, it is obvious that in
order to extract information of the averaged forcing field an enormous
number of realizations is necessary. This numerical procedure now
complies with the path integral formulation for
the observable $\mathcal{O}(u)=\langle \delta(u_x(0,0)=a)\rangle$
\begin{equation}
\langle \mathcal{O}(u)\rangle = \int \mathcal{D}f \, \mathcal{O}(u)
\delta(u_t+uu_x-\nu u_{xx} - \phi) \mbox{e}^{-(\phi,\chi^{-1} \phi)/2}
\end{equation}
which is the starting point for the Martin-Siggia-Rose formulation. Thus, for
sufficiently strong velocity gradients $u_x(0,0)=a$, the important question and
conjecture is whether the averaged solutions $\langle u_{\text{shifted}}(t,x)\rangle$
and $\langle \phi_{\text{shifted}}(t,x)\rangle$ coincide with the instanton solution of
(\ref{instanton}). Especially if this conjecture is true, then
the averaged optimal force $\langle \phi_{\text{shifted}}(t,x)\rangle$ should coincide with the right-hand side of equation
(\ref{instantonu})
\begin{equation}
\langle \phi_{\text{shifted}}(t,x)\rangle = -i \int \chi(x-x')p(x',t)dx'
\end{equation}
where the auxiliary field $p(t,x)$ is obtained from the direct Chernykh-Stepanov
algorithm.
Fig. (\ref{fig:comparisonInstantons}) shows the filtered field $\langle
u_{\text{shifted}}(t,x)\rangle$ (top) and the instanton field $u(t,x)$ (bottom)
at the final time $t=0$ and at an earlier time $t=-1.75$ showing the instanton
in a different phase (see also the sketch of the instanton phases in Fig. 8 in
\cite{chernykh-stepanov:2001}). The agreement is remarkable. Especially the
center region is precisely reproduced by the stochastic simulation, while the
sides are less pronounced. In order to get a complete overview of the time
history of the instanton and the filtered field, Fig. (\ref{fig:comparisonInstantonsContour}) depicts a contour plot
of the whole space-time domain. Although the filtered field shows a slightly 
shorter extent in time, the congruence is clearly visible.

\begin{figure}[t]
\begin{center}
\includegraphics[width=\columnwidth]{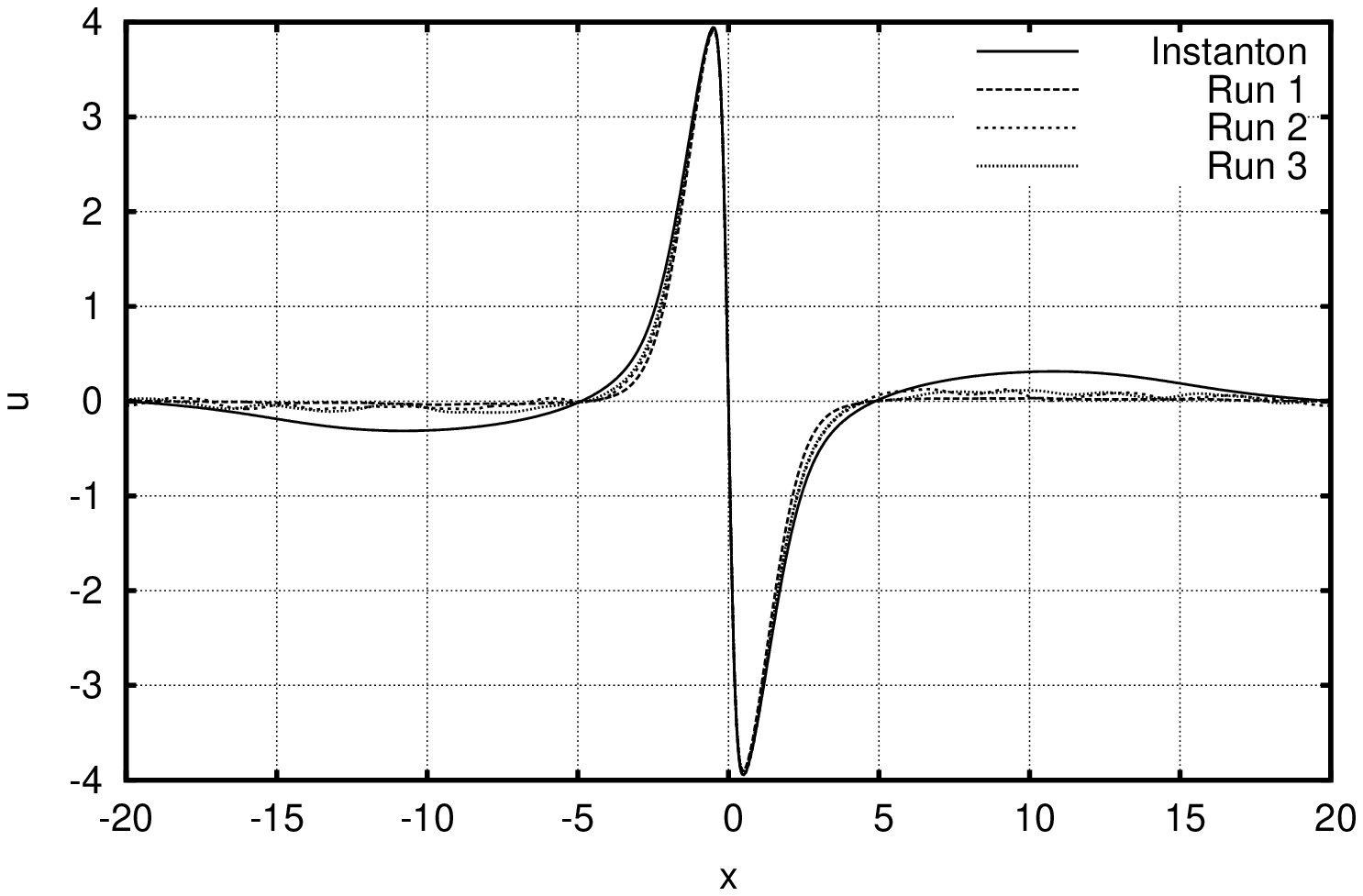}\\
\includegraphics[width=\columnwidth]{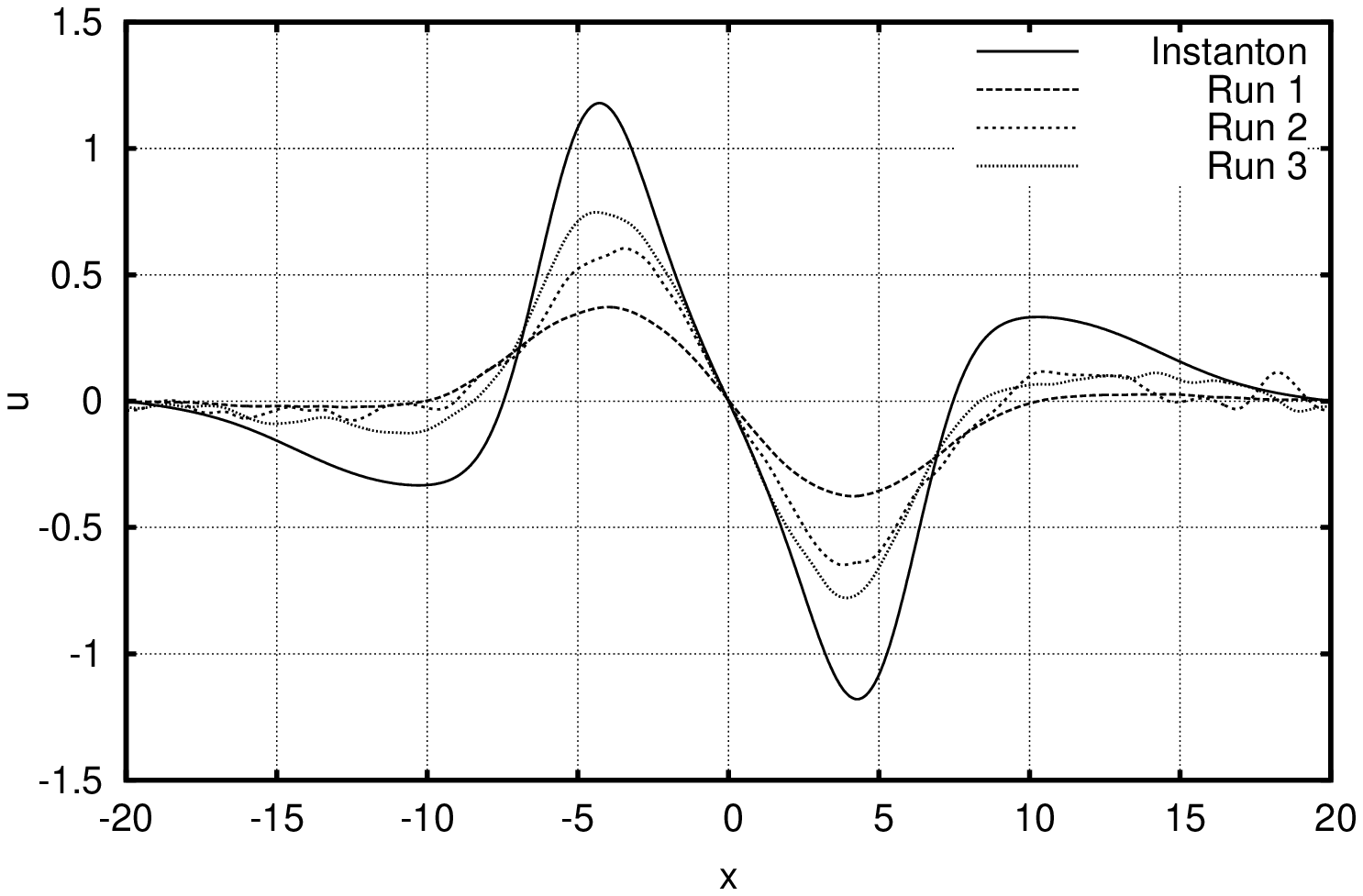}
\end{center}
\caption{Comparison of the instanton field $u(t,x)$ (solid) to
  stochastic simulations with varying hit percentages ($\approx 10\%$
  (dashes), $\approx 0.5\%$ (dots), $\approx 0.05\%$ (small dots)) for
  $t=0$ (top) and $t=-1.75$ (bottom). Agreement with the instanton
  approach increases with decreasing hit percentage.
  \label{fig:rareness}}
\end{figure}

The rareness of the filtered events has a strong impact on the
agreement between the instanton approximation and the full stochastic
simulation. In order to demonstrate the varying resemblance to the
instanton approximation, we alter the probability of reaching a
prescribed velocity gradient by changing the kinematic viscosity
$\nu$. Fig.~(\ref{fig:rareness}) shows the filtered field $\langle
u_{\text{shifted}}(t,x) \rangle$ and the instanton field $u(t,x)$
for three different hit percentages. As the rareness of the event
increases, accordance with the instanton grows
considerably. Especially the velocity gradient in the origin is only
reproduced when the events are rare. Notably this effect does not
depend on the Reynolds number or shock strength, but on the scarcity
of the event alone.

\begin{figure}[t]
\begin{center}
\includegraphics[width=\columnwidth]{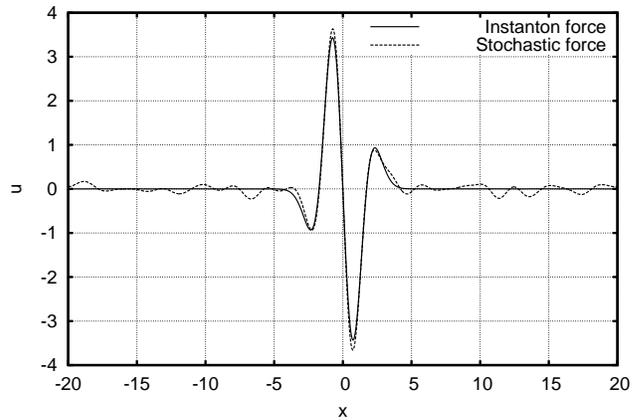}
\end{center}
\caption{The filtered force field $\langle
  \phi_{\text{shifted}}(t,x)\rangle$ (dashed) and the analytical force
  field $4 \nu^2 \mathcal{F} \chi'(x)$ (solid) at time $t=0$.
\label{fig:forcefield}}
\end{figure}

An additional feature of this filtering approach is that not only the instanton
velocity field could be extracted but also the time history of the auxiliary
field $p(t,x)$ and of the optimal force field. At time $t=0$, the auxiliary
field is given by its initial condition $p(t=0,x) = -4i\nu^2{\mathcal{F}}
\delta'(x)$ and produces the force term $4\nu^2 \mathcal{F} \chi'(x)$ on the
right-hand side of eqn. (\ref{instantonu}). A comparison of this term with the
filtered force field $\langle \phi_{\text{shifted}}(t,x)\rangle$
is depicted in Fig. (\ref{fig:forcefield}), which shows a remarkable agreement. \\

\noindent {\it Conclusions and Outlook}
In this Letter we studied the question whether the instanton solution for
Burgers turbulence is ``real'', e.g. can be observed in stochastically driven
simulations. The positive answer to this question is remarkable since we observe
the instanton for a moderate Reynolds number and thus for moderate (and not
extreme) values of the velocity gradient. In principle, this filtering method
also allows a further study of moderately scarce events to determine where
fluctuations around the instanton appear, how they look like and how they modify
the action integral and thus the PDF. Our findings could also open the door to
the issue why the PDF for the very left tail of velocity gradients could not be
observed in high resolution numerical experiments of Gotoh~\cite{gotoh-1999}.
Although this is out of the scope of the
present Letter, work in this direction is in progress. \\

\noindent {\bf Acknowledgment}
We acknowledge stimulating discussion with M. Polyakov, R.
Friedrich and M. Wilczek. Especially, we would like to thank M. Rieke for
sharing his CUDA programming expertise. This work benefited from partial support
through DFG-FOR1048 and the NSF grants DMS-0807396, DMS-1108780, and
CNS-0855217. Numerical simulations where conducted on the CUDA-Cluster of the
Research Department Plasma Physics at the Ruhr-University Bochum and the
GPU-Cluster at the High Performance Computing Center of the City University of
New York.

This paper is dedicated in memoriam of Professor Rudolf Friedrich (\textdied 16.8.2012) .

\bibliography{bib}

\end{document}